\documentstyle[prb,aps,twocolumn]{revtex}  
\input epsf
\epsfverbosetrue
\tighten
\begin{document}
\draft
\twocolumn[\hsize\textwidth\columnwidth\hsize\csname @twocolumnfalse\endcsname

\title{Time evolution of tetragonal-orthorhombic ferroelastics}
 
\author{S. H. Curnoe\cite{Stephanie} and A. E. Jacobs\cite{Allan}}

\address{Department of Physics, University of Toronto,
Toronto, Ontario, Canada M5S 1A7}
 
\date{\today}
 
\maketitle
 
\begin{abstract}
We study numerically the time evolution of two-dimensional (2D) domain 
patterns in proper tetragonal-orthorhombic (T-O) ferroelastics. 
Our results, found by solving equations of motion derived from classical 
elasticity theory, disagree with those found by other methods. 
We study first the growth of the 2D nucleus resulting from homogeneous 
nucleation events. 
The later shape of the nucleus is largely independent of how it was nucleated. 
In soft systems, the nucleus forms a flower-like pattern. 
In stiff systems, which seem to be more realistic, it forms an X shape with 
twinned arms in the 110 and $\bar110$ directions. 
Second, we study the relaxation that follows completion of the phase 
transition; 
at these times, the T phase has disappeared and both O variants are present, 
segregated into domains separated by domain walls. 
We observe a variety of coarsening mechanisms, most of them counterintuitive. 

\end{abstract}
 
] 
\narrowtext
\tightenlines

\section{Introduction}
Ferroelastics \cite{aizu,salje93} are crystalline solids that undergo a 
shape-changing phase transition, usually first-order, to a state of lower 
symmetry with decreasing temperature $T$. 
A prominent example is the tetragonal-orthorhombic (T-O) ferroelastic 
YBa$_2$Cu$_3$O$_7$.\cite{salje93,king} 
At the transition temperature $T_c$, the unit cell of the parent (high-$T$) 
phase distorts spontaneously in one of several equivalent directions. 
Each of these degenerate distortions corresponds to a differently oriented 
{\it variant} of the product (low-$T$) phase. 
Below $T_c$, all variants are usually present, separated from each other by 
domain walls with preferred orientations. 
Domain patterns in ferroelastics differ greatly from those in ferromagnets, 
gainsaying the analogy responsible for the very name ferroelastic \cite{aizu} 
and confounding intuition based on conventional order-parameter systems. 
The difference from these other systems is that the strains are not 
independent order parameters but rather are linked by compatibility relations. 

The theory of proper ferroelastics (where the strain is the primary order 
parameter) extends the classical theory of elastic continua by adding 
higher-order terms in the strains and also derivatives of the strains. 
This strain-only theory was first used \cite{falk} in one dimension (1D); 
its first important result was a remarkable solution \cite{barsch} for the 
twin wall of cubic-tetragonal (C-T) materials. 
It has since been used to study various aspects of 
(a) T-O materials,
\cite{jacobs85,kartha,jacobs95,kerr99,onuki99,shenoy99,jacobs00}
(b) the 1D problem,\cite{reid94} 
(c) cubic-tetragonal (C-T) materials,\cite{luskin,ras00,curnoe1} and 
(d) hexagonal-orthorhombic (H-O) and related materials.\cite{curnoe2} 
Although the strain-only theory applies strictly only to proper 
ferroelastics, it has nevertheless succeeded in explaining a wide variety of 
domain patterns also in improper T-O \cite{jacobs00} and H-O \cite{curnoe2} 
materials. 
A much larger literature (examples are 
Refs. \onlinecite{semen91,semen93,brat96,saxena97,yama98,yama00,wen99}) 
includes order parameters in addition to the strains or applies more 
phenomenological approaches. 

The basic formalism for describing the time evolution of proper ferroelastics, 
though known for a century,\cite{landau} has been used only infrequently, to 
study 1D,\cite{reid94} C-T \cite{luskin} and H-O \cite{curnoe2,reid97} 
systems. 
Fundamentally different dynamical schemes were used in the strain-only 
theories of Refs. 
\onlinecite{kartha,kerr99,onuki99,shenoy99,ras00,saxena97}, often only 
as a tool to find static structures. 

The following presents the first application of the classical equations of 
motion \cite{landau} to the dynamics of proper T-O ferroelastics. 
The study was motivated in part by the electron-microscopy results 
\cite{salje93,king} available for YBa$_2$Cu$_3$O$_7$; 
this is an improper material (the orthorhombic distortion is a secondary 
effect of the oxygen ordering), but the success of the static strain-only 
theory \cite{jacobs00} for YBa$_2$Cu$_3$O$_7$ warrants an extension to the 
dynamics. 
Computational resources allow us to consider only 2D structures, with 
possible application to thin films, particularly to the patterns of 
Refs. \onlinecite{salje93,king}. 

The paper is organised as follows. 
Section II gives the expression for the T-O strain energy in 2D and then 
finds the equations of motion. 
We distinguish between soft and stiff systems according to the energy cost 
for wall directions off optimal. 
The strain-only theory predicts softening with decreasing $T$, perhaps with 
observable consequences. 
Section III applies this formalism to investigate the growth of O nuclei from 
the supercooled T phase. 
We find that the developed nuclei are largely independent of the nucleation 
mechanism; 
in both soft and stiff systems they differ markedly from the nuclei in other 
theories.\cite{shenoy99,yama98} 
In soft systems, nuclei are flower-like; 
they simply expand without generating much additional structure. 
In stiff systems, they form an X shape with twinned arms in the major growth 
directions (110 and $\bar110$); 
additional structure forms near the centre and propagates outward along the 
arms. 
Section IV examines the coarsening mechanisms that follow completion of the 
phase transition, including domain-wall merges, formation and disappearance 
of island domains, rank formation of ribbon tips and their coordinated 
retraction, and tip splitting (in stiff systems). 
Section V provides a summary and proposals for further investigations. 

\section{Equations of Motion}
\subsection{Expansion of the strain energy}
The energy of proper ferroelastics is expressed solely in terms of the strains. 
These are combinations of derivatives of the displacement ${\bf u}({\bf x})$ 
of a material point from its position ${\bf x}$ in the high-$T$ symmetric 
phase.
We discuss only structures uniform in the tetragonal fourfold ($x_3$) 
direction. 
We define the three strains in 2D by 
\renewcommand{\theequation}{1\alph{equation}}
\setcounter{equation}{0}
\begin{eqnarray}
e_1 & = &(\eta_{11} + \eta_{22})/\sqrt{2} \ , \\
e_2 & = &(\eta_{11} - \eta_{22})/\sqrt{2} \ ,  \label{e2}\\
e_3 & = &(\eta_{12} + \eta_{21})/2 \ ,
\end{eqnarray}
\renewcommand{\theequation}{\arabic{equation}}
\setcounter{equation}{1}
where the components of the strain tensor $\eta$ are 
\begin{equation}
\eta_{ij}=\frac{1}{2}\left(u_{i,j}+u_{j,i}+u_{k,i}u_{k,j}\right) \ ;
\label{tensor}
\end{equation}
here $u_{i,j}=\partial u_i/\partial x_j=\partial_j u_i$ and repeated indices 
are summed. 
All three strains vanish in the T phase. 
The deviatoric strain $e_2$ is the primary order parameter of the T-O 
transformation. 
In the lowest-energy product state, $e_2$ takes one of two degenerate values  
$\pm e_{20}$ corresponding to a stretch in either the $x_1$ or the $x_2$ 
direction. 
The dilatational and shear strains $e_1$ and $e_3$ vanish for these two 
states, and also for twin bands \cite{jacobs85}, but not for the complex 
domain patterns formed by colliding bands. 

At this stage in the theory of ferroelastics, one wants to examine the 
simplest possible form for the energy density ${\cal F}$, to include only 
those terms required by symmetry, for stability, and to explain experiment. 
We start from the expression 
\begin{eqnarray}
{\cal F} =&& \frac{1}{2}\left[A_1e_1^2 + A_2\left(T\right)e_2^2 
         + A_3e_3^2\right] +\frac{B_2}{4}e_2^4 +\frac{C_2}{6}e_2^6  
\nonumber \\
  &+& \frac{d_2}{2}\left(\nabla e_2\right)^2 \ ;
\label{free}
\end{eqnarray}
all terms are invariant under the symmetry operations of the T group. 
The dilatational, deviatoric and shear stiffnesses $A_1$, $A_2(T)$ and $A_3$ 
in the first term are related to the elastic constants. 
Stability requires $A_1\geq0$ and $A_3\geq0$. 
But $A_2(T)$ softens with decreasing $T$, as $A_2(T)=a(T-T_0)$, and the T 
phase is unstable for $T<T_0$. 
To describe the phase transition, we need the terms in $e_2^4$ and $e_2^6$; 
we assume a first-order transition ($B_2<0$), and so $C_2>0$ for stability. 
At high $T$, namely $A_2(T)> B_2^2/4C_2$, only the T minimum exists. 
At lower $T$, two O minima occur at $e_2 = \pm e_{20}(T)$, where
\begin{equation}
e_{20}(T)  = \left[\left(-B_2 + \sqrt{B_2^2-4A_2(T)C_2}\right)
/\left(2C_2\right)\right]^{1/2}\ .
\end{equation}
At the transition temperature $T_c$, found from $A_2(T_c) = 3 B_2^2/16C_2$, 
the three minima $e_2 = 0$, $\pm e_{20}(T_c)$ are degenerate; 
here $e_{20}(T_c) = \sqrt{-3B_2/4C_2}$. 
Finally, the gradient term is responsible for the wall energy; 
the other derivative invariants \cite{jacobs92} are unimportant,
\cite{kartha,jacobs00,jacobs92} largely because the primary physical spatial 
dependence is in $e_2$. 

The parameters of the theory are not well known for any material. 
To reduce the number of unknown parameters, and possibly obtain a universal 
theory that applies qualitatively to many materials, we transform variables 
by
\begin{eqnarray}
\label{scalea}
e_j &      \to & \left[e_{20}(T_c)\times 10^3\right]           \ e_j     \ ,\\
\label{scaleb}
x_i &      \to & \sqrt{d_2/A_2(T_c)}                           \ x_i     \ ,\\
\label{scalec}
{\cal F} & \to & A_2(T_c)\left[e_{20}(T_c)\times 10^3\right]^2 \ {\cal F}\ ;
\end{eqnarray}
also, we define the dimensionless temperature 
$\tau = A_2(T)/A_2(T_c)=(T-T_0)/(T_c-T_0)$ and 
dimensionless stiffness parameters 
$\zeta_1 = A_1/A_2(T_c)$ and $\zeta_3 = A_3/A_2(T_c)$. 
The scale factor in Equation (\ref{scalea}) is chosen so that the deviatoric 
strain at $T_c$ is $10^{-3}$, an arbitrary value; 
the hidden but necessary assumption here is that the strains are small and so 
the nonlinear term in Equation (\ref{tensor}) can be neglected. 
The energy density in terms of the new variables is
\begin{equation}
{\cal F} = \frac{1}{2}\left(\zeta_1 e_1^2 + \tau e_2^2 + \zeta_3 e_3^2\right) 
+\frac{b}{4}e_2^4 +\frac{c}{6}e_2^6 + \frac{1}{2}\left(\nabla e_2\right)^2
\label{free2}
\end{equation}
where $b = -4\times10^{6}$ and $c=3\times10^{12}$. 
If temperatures near $T_c$ are accessible, the three parameters in 
Equation (\ref{free2}) can be determined from the elastic constants just 
above $T_c$, the strain $e_2$ at $T_c$ and the $T$-dependence of $e_2$. 
For YBa$_2$Cu$_3$O$_7$, typical values at low $T$ are \cite{cai98} an 
orthorhombic distortion of $2(b-a)/(a+b)=0.017$ (giving $e_{20}=0.012$), and 
a wall width of $\approx1.3\;$nm. 

Static structures predicted by Equations (\ref{free}) (or (\ref{free2})) 
are discussed in Refs. \onlinecite{jacobs85,jacobs95,jacobs00}. 
Domain walls have lowest energy ($e_1$ and $e_3$ are zero) when in the T 
110 and $\bar110$ planes. 
The walls link the variants but also rotate them by an angle proportional to 
$e_2$. 
The rotation, which has no counterpart in conventional order-parameter 
systems, gives rise to unusual effects when orthogonal walls collide; 
for example, the visual wall length increases in the collision region, due to 
variant narrowing \cite{jacobs00} resulting from formation of a disclination. 

Different structures are found in soft or stiff systems, depending on whether 
the energy cost is small or large for wall directions off the optimal 110 and 
$\bar110$ planes. 
The relevant parameters are the ratios $\zeta_1/\zeta_2$ and $\zeta_3/\zeta_2$ 
of the dilatational and shear stiffnesses to the deviatoric stiffness 
$\zeta_2=\tau + 3b e_{20}^2 + 5c e_{20}^4$. 
The energy cost increases with both ratios, though more strongly with 
$\zeta_1/\zeta_2$ it seems. 
Strangely, systems soften with decreasing $T$, because $\zeta_2$ increases 
($\zeta_2\to4$ as $\tau\to1$ from below, with $\zeta_2=238$ at $\tau=-50$ for 
example). 
If a system is moderately stiff just below $T_c$, then features like split 
tips characteristic of stiff systems may disappear on cooling, provided that 
low enough temperatures are accessible and that the relaxation is not too 
sluggish. 

\subsection{Time evolution}
The Lagrangian density is
\begin{equation}
{\cal L} = {\cal T}-{\cal V} = \frac{1}{2}\rho(\dot u_i)^2 - {\cal F}
\end{equation}
where ${\cal F}$ is the strain-energy density. 
To represent the nonconservative forces in the system, we use a Rayleigh 
dissipative function,\cite{landau} with density 
\begin{equation}
{\cal R} = \frac{1}{2}\left(  A'_1\dot{e}_1^2 + A'_2\dot{e}_2^2 
                            + A'_3\dot{e}_3^2 \right) \ ;
\label{Psieq}
\end{equation}
here $\dot{e}_j=\partial e_j/\partial t$. 
This form respects the symmetry of the T phase; 
it assumes evolution without plastic flow. 
The important point is that Equation (\ref{Psieq}) leads to dissipative 
forces that 
are functions of the spatial derivatives of the velocity, as one would 
expect, since uniform motion of the material cannot dissipate energy. 
Then the equations of motion are \cite{landau} 
\begin{equation}
\rho \ddot{u}_{i} - \sigma'_{ik,k} - \sigma_{ik.k} = 0
\label{mot}
\end{equation}
where
\begin{eqnarray}
\sigma'_{ki} & = & \frac{\partial {\cal R}}{\partial \dot{u}_{i,k}} \ ,\\
\sigma _{ki} & = & \frac{\partial {\cal F}}{\partial u_{i,k}} \ .
\end{eqnarray}

We assume that the dissipative term is much larger than the inertial term. 
This approximation fails however at small wavenumber, as discussed for 
example in Ref. \onlinecite{reid94}. 
In particular, the zeroth Fourier component should be considered separately
since the last two terms of Equation (\ref{mot}) are then zero; 
then the inertial term tells us that the motion is uniform, determined by the 
initial value.  
Without the inertial term, Equation (\ref{mot}) simplifies to 
\begin{equation}
\sigma'_{ik,k} =  - \sigma_{ik,k} \ .
\label{mot2}
\end{equation}
The summation on the index $k$ prevents integration of these equations, 
except in 1D. 
In 1D, the constant of integration is crucial, for it represents 
external forces applied to the boundary that may hold the system in a 
static configuration that is not necessarily the unconstrained minimum 
of the strain energy.  

The equations of motion (\ref{mot2}) in terms of the strains are 
\renewcommand{\theequation}{15\alph{equation}}
\setcounter{equation}{0}
\begin{eqnarray}
A_1'\dot{e}_{1,1} + A_2'\dot{e}_{2,1} +\frac{A_3'}{\sqrt{2}}\dot{e}_{3,2}
&=& -(G_{1,1}+G_{2,1} + G_{3,2}) \label{eofm1} \ ,\\
A_1'\dot{e}_{1,2} - A_2'\dot{e}_{2,2} +\frac{A_3'}{\sqrt{2}}\dot{e}_{3,1}
&=& -(G_{1,2}-G_{2,2} + G_{3,1})  \label{eofm2}\ ,
\end{eqnarray}
\renewcommand{\theequation}{\arabic{equation}}
\setcounter{equation}{15}
where $G_i=\delta {\cal F}/\delta e_i$ and the individual functionals are 
\renewcommand{\theequation}{16\alph{equation}}
\setcounter{equation}{0}
\begin{eqnarray}
G_1 & = & A_1 e_1                                           \ ,\\
G_2 & = & A_2 e_2 +B_2 e_2^3 + C_2 e_2^5 - d_2 \nabla^2 e_2 \ ,\\
G_3 & = & A_3 e_3/\sqrt{2}                                  \ . 
\label{mot3}
\end{eqnarray}
\renewcommand{\theequation}{\arabic{equation}}
\setcounter{equation}{16}

We emphasize that our equations of motion (15) are not those of 
time-dependent Ginzburg-Landau (TDGL) theory. 
Schematically, the latter are 
\begin{equation}
\dot{e_i} \propto - \delta{\cal F}/\delta e_i \ ,
\label{tdgl}
\end{equation}
with a nonlocal expression \cite{nonlocal} for the density ${\cal F}$; 
the major difference is the additional space derivatives on both sides of 
Equation (15). 
Equation (\ref{tdgl}) has much intuitive appeal, not least because it 
continues the analogy with ferromagnets. 
Nevertheless, it cannot be correct in principle, and in fact its predictions 
disagree with those of Equation (15). 
We illustrate the point by considering a material with short-range internal 
forces, uniformly stretched by external forces applied at the ends. 
When the latter are abruptly released, relaxation begins at the ends and 
propagates inward, taking a finite amount of time to reach any point in the 
bulk; 
the ions (except those near the ends) feel equal but opposite forces from 
their neighbours until the disturbances reach their vicinity. 
Equations (15) have the correct behaviour, whereas Equation (\ref{tdgl}) 
predicts instantaneous response. 

Equations (15) differ also from the dynamics 
\begin{equation}
{\dot u}_i \propto -\delta {\cal F}/\delta u_i 
\label{udot}
\end{equation}
of Refs. \onlinecite{kartha} and \onlinecite{onuki99}, the former at $T=0$. 
Not having examined physical settings comparable to those where Equation 
(\ref{udot}) was used, we cannot compare its results with those of Equation 
(15). 
The right-hand side of Equation (\ref{udot}) agrees that of 
Equation (\ref{mot2}); 
but the left-hand side, a dissipative force proportional to the velocity, 
cannot be correct in principle. 

From Equations (15), the equations of motion for the two components of 
{\bf u} are 
\begin{eqnarray}
&&\left(\begin{array}{cc}
(A_1'+A_2')\partial_1^2 +A_3'\partial_2^2/2 &
(A_1'-A_2'+A_3'/2) \partial_1\partial_2 \\
(A_1'-A_2'+A_3'/2) \partial_1\partial_2  &
(A_1'+A_2')\partial_2^2 +A_3'\partial_1^2/2 \end{array}\right)
\left(\begin{array}{c} \dot{u}_1 \\ \dot{u}_2 \end{array} \right) \nonumber \\
&& \hspace{.4in}=
-2\left(\begin{array}{c} G_{1,1}+G_{2,1}+G_{3,2}\\G_{1,2}-G_{2,2}
+G_{3,1} \end{array} \right) \ .
\label{65}
\end{eqnarray}
By solving these equations, we satisfy automatically the 2D compatibility 
relation 
\begin{equation}
\nabla^2 e_1 -(\partial^2_1-\partial^2_2)e_2-\sqrt{8}\partial_1
\partial_2e_3=0 
\end{equation}
in the small-strain approximation. 
This necessary and sufficient requirement that the strains be derivable from 
the displacement can be obtained by starting from $u_{i,12} = u_{i,21}$. 

The three viscosity parameters $A_i^\prime$ are not known from experiment, 
though of course all must be $\geq0$; 
it is then reasonable to consider the simplest possible theory. 
In choosing parameter sets, we should avoid those that give a vanishing 
determinant 
\begin{eqnarray}
\mbox{Det} =& & \frac{1}{2}(A_1'+A_2')A_3'(\partial_1^4+\partial_2^4)
\nonumber \\
&+&\left[4 A_1'A_2'-(A_1'-A_2')A_3'\right]\partial_1^2\partial_2^2
\end{eqnarray}
of the coefficients on the left-hand side of Equation (\ref{65}); 
inspection shows that only one of the $A_i^\prime$ can vanish. 
Other cases of interest are those for which the determinant factors, 
{\it i.e.} 
\begin{equation}
4 A_1'A_2'-(A_1'-A_2')A_3' = \pm (A_1'+A_2')A_3' \ , 
\end{equation}
giving three possibilities: 
(a) $A_1'= 0$, 
(b) $A_3'  = 2A_2'$, and 
(c) $A_2'=0; $
a fourth, namely $A_3'=- 2 A_1'$ fails on grounds discussed above. 
Since $e_2$ is the primary order parameter, we should keep $A_2'$; 
the time scale is then adjusted so that $A_2^\prime=1$. 
The choices $A_3'=2A_2'$ (the isotropic case) and $A_1'=0$ are convenient, 
for then the left-hand sides of Equation (\ref{65}) decouple. 
We verified that taking $A_1^\prime=1$ {\it vs.} $A_1'=0$ has little effect 
during the evolution; 
the fully relaxed configurations can differ however. 

We imposed periodic boundary conditions on the displacement {\bf u}, thereby 
forcing domain walls into the systems; 
the equilibrium states are a single twin band, optimally with only a pair 
of walls. 
We solved Equations (\ref{65}) using a finite-difference, 
fast-Fourier-transform method. 
At the beginning of each time step, the displacement field was known at 
each point of the space grid. 
Finite-difference approximations (centered on a $5\times5$ grid) were used to 
compute the derivatives and so to obtain the right-hand sides in real space. 
The latter were then Fourier transformed. 
The Fourier components on the left-hand sides were found using the same 
finite-difference approximations and then advanced in time using the Euler 
method (with time step $10^{-5}$ or so). 
The results were then Fourier transformed back to real space to begin the 
next step. 

\section{T-O nucleus in two dimensions}

This section studies the nucleus resulting from perturbing the supercooled 
T phase in various ways. 
All results are for a grid of $512\times512$ points, with step size 0.4. 

We first present results obtained by displacing a single point off a 
high-symmetry direction. 
Figure 1 shows snapshots for $\tau=-50$ ($\zeta_2=238$) and for four sets of 
values of $\zeta_1$ and $\zeta_3$, all at time $t=0.18$ after identical 
nucleation events; 
the viscosity parameters are $A_1'=0$, $A_2'=1$ and $A_3'=2$. 
Figure 2 shows snapshots of the same systems at the later time $t=0.24$. 
Very little is known about the relative importance of the stiffnesses 
$\zeta_1$ and $\zeta_3$ and so we investigated some extreme cases; 
we find stronger dependence on $\zeta_1$ than on $\zeta_3$. 
Parts (a) and (b) of Figures 1 and 2 show soft systems ($\zeta_1=1$), with 
$\zeta_3=1$ and 1000 respectively, whereas parts (c) and (d) show 
moderately stiff systems ($\zeta_1=1000$), again with $\zeta_3=1$ and 1000 
respectively. 
The important point is that the nucleus has very different shapes in soft and 
stiff systems; 
one notes also the more rapid growth in the latter. 

\noindent 
In the soft systems, the domain walls lie off the optimal directions; 
the nucleus retains its disk shape as it expands. 

\noindent 
In the stiff systems, the domain walls are much closer to the optimal 
orientations. 
The nucleus has a striking X shape with arms in the 110 and $\bar110$ 
directions; 
growth transverse to the arms results from the appearance of new variants 
near the nucleation site and their subsequent growth along the arms. 

Other sets of simulations started from point displacements in high-symmetry 
directions (100 and 110), and others from displacements of small areas. 
Every soft system gave a disk with eight or more distinct domains emanating 
from the disturbed area; 
every stiff system gave the X-shaped nucleus. 

Yet more sets of simulations started with 
disk-shaped regions containing several parallel 
stripes in one direction, this in a bid to approximate the nucleus reported 
in Ref. \onlinecite{shenoy99}. 
These attempts gave nuclei much like those from a point perturbation. 
Parts (a1) and (a2) of Figure 3 show that a soft system evolves toward the 
flower-like patterns in parts (a) and (b) of Figures 1 and 2. 
Parts (b1) and (b2) of Figure 3 show the evolution of a stiff system; 
the overall size of the figures is identical to those for the soft system, 
but the area of the starting configuration is about one-fourth that in 
parts (a1) and (a2). 
Fast growth occurs parallel to the starting walls, but twinned jets shoot out 
in the transverse direction, thereby evolving the system toward the X shape 
in parts (c) and (d) of Figures 1 and 2.   
The two sets of jets are more asymmetric here, because the rapid longitudinal 
growth exaggerates the greater asymmetry in the starting configuration. 
Nevertheless, it is clear that even this starting configuration is also 
unstable toward the formation of perpendicular jets and evolution to the 
X shape.  

Simulations at other temperatures (between $\tau=-100$ and $\tau=-5$) gave 
results qualitatively similar to those described in Figures 1 to 3; 
the major difference is that the nucleus grows more slowly at higher $T$, 
as expected. 
The important point is that the flower/X shapes were found for soft/stiff 
systems at all $T$. 
We were unable to nucleate the low-$T$ phase above $\tau=-5$ (well below the 
stability limit $\tau=0$ of the T phase) and so we could not examine the 
parameter set of Ref. \onlinecite{shenoy99}. 

Because the gross features are independent of the starting configurations 
and temperature, we believe that we have found the nucleus of the T-O 
transformation in 2D, with possible application to thin films. 
It is reasonable to expect that $x$-$y$ cuts through the 3D T-O nucleus will 
resemble our 2D nucleus. 

None of our simulations (with any starting configuration, with either soft or 
stiff parameters, at any temperature) gave a nucleus resembling that found 
using TDGL theory in the strains. 
The 2D T-O nucleus of Ref. \onlinecite{shenoy99} accords with one's 
intuition based on conventional systems. 
It is compact, elliptical in shape (with axes along the 110 and $\bar110$ 
directions), and internally twinned (with walls parallel to the major axis); 
the twinning generates both positive and negative displacements which 
largely cancel overall. 
Transverse growth occurs by adding walls and variants, whereas existing 
variants grow only longitudinally. 
Although other aspects are different (Ref. \onlinecite{shenoy99} studied 
soft systems, used a somewhat different strain-energy functional, and worked 
at higher $T$, namely $\tau=0.3$), it is likely that the different results 
reflect the different dynamics. 

None of our simulations gave a nucleus like that in the more phenomenological 
study of Ref. \onlinecite{yama98}, namely growth to an untwinned square 
which then flowers. 

The only previous use of the equations of motion (\ref{mot2}) to examine 
nucleation was in a study of H-O ferroelastics;\cite{curnoe2}
these systems are dominated by disclinations. 
In soft systems the nucleus is flower-like, as in T-O systems, but has 12 
arms; 
in stiff systems it branches early in the growth, without forming the long 
arms seen above in T-O systems. 

\section{Coarsening}

This section studies the coarsening phenomena that occur after completion of 
the phase transition. 
The interest lies in the unconventional behaviour relative to that observed 
in order-parameter systems. 
Simulations started from systems with orthogonal twin bands, relaxed 
internally but not in the collision regions. 
The initial relaxation from these artificial high-energy configurations is 
rapid and of no interest; 
we present results at later times, but well before equilibrium is reached. 

Figure 4 shows four pairs of snapshots. 

\noindent Parts (a) to (c) are for soft systems with different initial 
conditions, all with parameters $A_1'=A_2'=A_3'=1$, $\zeta_1=\zeta_3=10$, 
and $\tau=-50$; 
the times between the pairs are 0.5, 0.5 and 1.0 respectively. 

\noindent In part (a), the island at the centre vanishes, but other islands 
form as some narrow domains pinch off and retract. 

\noindent In part (b), one tip retracts to form rank with its neighbour; 
at the lower right, other tips retract in unison, keeping the rank. 

\noindent In part (c), coarsening occurs by different kinds of coordinated 
events; domain merges parallel to the smaller-scale patterns occur at the top 
left and perpendicular at the bottom right. 

\noindent Part (d) corresponds to a stiffer system, with parameters 
$A_1'=A_2'=1$, $A_3'=2$, $\zeta_1=\zeta_3=500$, and $\tau=-100$ 
($\zeta_2=452$); the time difference is 0.6. 
The patterns are strikingly similar to Figures 7.9 and 7.17(b) of 
Ref. \onlinecite{salje93} and to a lesser extent Figure 2(b) of Ref. 3(b). 
One sees the formation of a split tip and also the counterintuitive variant 
narrowing and wall wobbling found in the static theory.\cite{jacobs00}
Related theories of needle twins and tip splitting are given in 
Refs. \onlinecite{kohn92} and \onlinecite{salje98}. 

The observation of tip splitting \cite{salje93,king} in YBa$_2$Cu$_3$O$_7$ 
suggests that this material is moderately stiff ($\zeta_1\agt\zeta_2$) at the 
temperatures investigated. 
Values of the elastic constants suggest that Fe-Pd alloys (cubic-tetragonal) 
are also moderately stiff.\cite{curnoe1}

These coarsening phenomena, like the nucleation phenomena reported in 
Section 3, confound intuition based on conventional order-parameter systems. 
The relaxation cannot be characterised by any simple rules; 
that is, the changes from one snapshot to the next cannot be predicted by 
inspection of the strain patterns alone. 
The visible domain-wall length often increases. 
The relaxation is nonlocal;\cite{nonlocal}
rapid changes occur in one part of the system while other parts, with no 
apparent major differences from the first, stay almost unchanged. 
The tendency is toward coarser patterns, but occasionally the topology 
becomes more complicated (as when islands form). 
The ribbons seldom retract immediately, even though retraction reduces the 
wall length. 
Particularly odd are the rank formation of tips and their linked withdrawal, 
the variant narrowing and the splitting of tips. 
Transverse wall motion occurs only locally, for example in the process of 
pinching off the other variant. 

Our simulations resemble in some respects those of Refs. 
\onlinecite{brat96,yama00}, and less those of Refs. 
\onlinecite{semen91,kerr99,shenoy99}. 

Coarsening mechanisms in simulations of H-O systems,\cite{curnoe2} also 
using Equations (\ref{mot2}), differ from those in Figure 4 (again due to 
the disclinations in H-O systems). 

\section{Summary}
We have derived general equations of motion for proper T-O ferroelastics
including inertia, dissipation and internal elastic stress.  
These equations, and more importantly their predictions, differ from those of 
all previous studies of proper T-O ferroelastics. 
We studied the growth of the O nucleus for both soft and stiff systems, in 2D. 
The soft system expands as a disk with time, while the stiff system assumes 
a characteristic X shape, with twinning along the arms. 
We studied also the coarsening mechanisms that relax the O phase toward 
local equilibrium, again in 2D. 
We observed the formation and disappearance of island domains, tip retraction 
and domain merging, both parallel and perpendicular to existing domain walls;  
in stiff systems we observed the formation of split tips.   
These mechanisms are likely not observable in proper ferroelastics, because 
the time scale is expected to be short; 
likely one can examine only patterns in quenched samples. 
Perhaps they are observable in improper systems, where the time scale may 
be longer; 
again, our strain-only theory does not apply in principle to improper 
ferroelastics, but it explains many puzzling features of patterns reported 
in Refs. \onlinecite{salje93} and \onlinecite{king}, and so perhaps it can 
shed light on the dynamics also. 

The above treatment should be extended to include thermal noise, first to 
examine the early stages of nucleus formation, and second to allow the 
system to surmount energy barriers. 
The tweed structure should be examined in the presence of noise, perhaps also 
with compositional fluctuations. 
The inertial term should be examined to determine whether it affects the 
dynamics significantly. 
The difficulty is to find realistic values of the viscosity parameters; 
one can easily be misled here. 

The primary need in the field is however {\it in situ} observations of the 
dynamics in T-O systems; 
these are difficult and correspondingly rare. 
The available studies \cite{salje98,vanten87} cannot decide the relative 
merits of the many theories. 

\acknowledgments 
This research was supported by the Natural Sciences and Engineering Research 
Council of Canada. 
We are grateful to E. K. H. Salje and R. C. Desai for discussions.

\begin{figure}[t]
\epsfysize=6.3in
\epsfbox[-20 00 620 580]{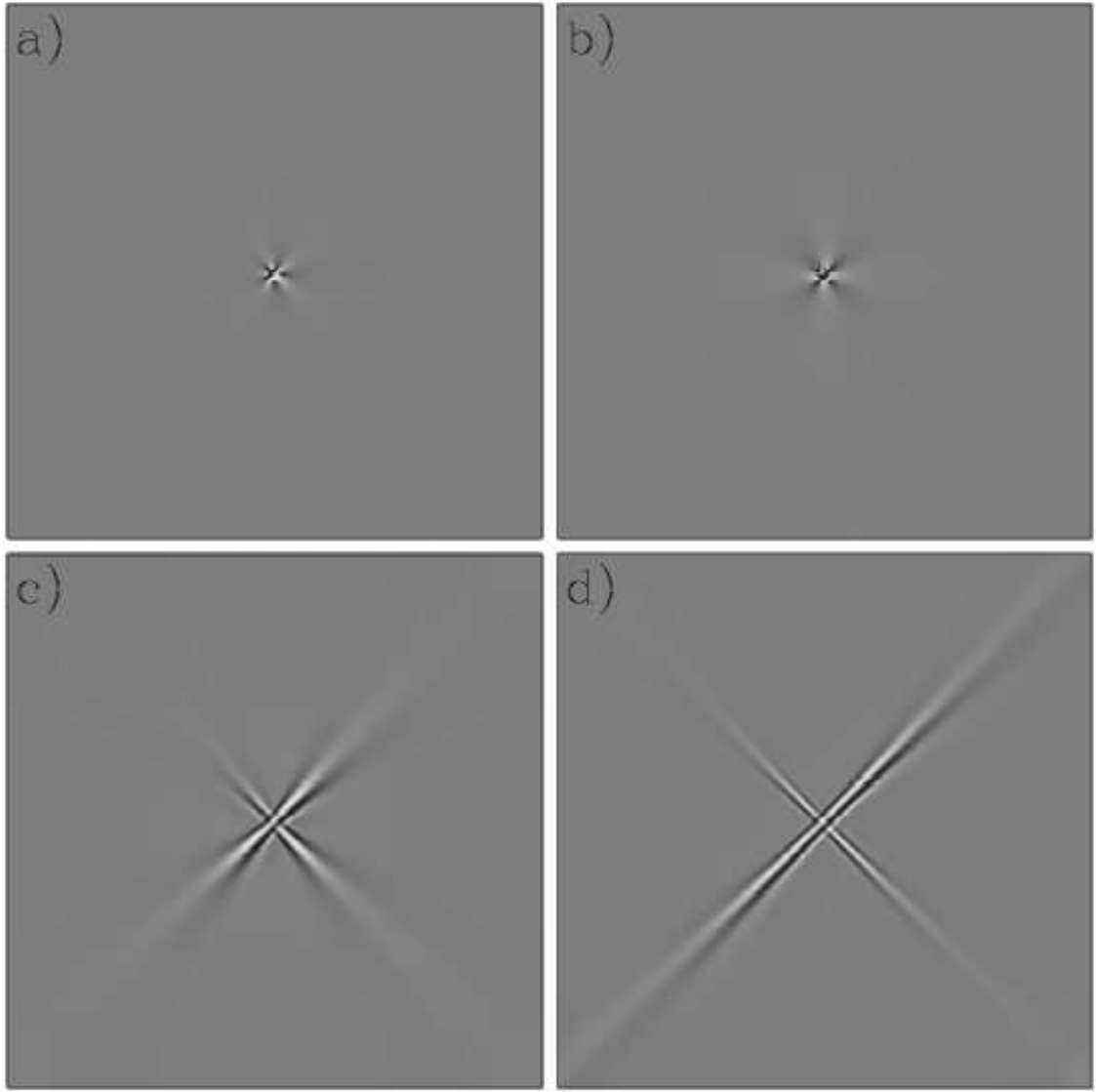}
\caption{Greyscale snapshots of orthorhombic (O) nuclei growing 
after identical perturbations of the supercooled tetragonal (T) phase. 
The two O variants are white and black, the T matrix grey. 
The four parts correspond to different choices for the dilatational 
and shear stiffnesses; the parameter values are given in the text. 
}
\end{figure}

\pagebreak
.

\pagebreak

\begin{figure}[t]  
\epsfysize=6.3in
\epsfbox[-20 00 620 580]{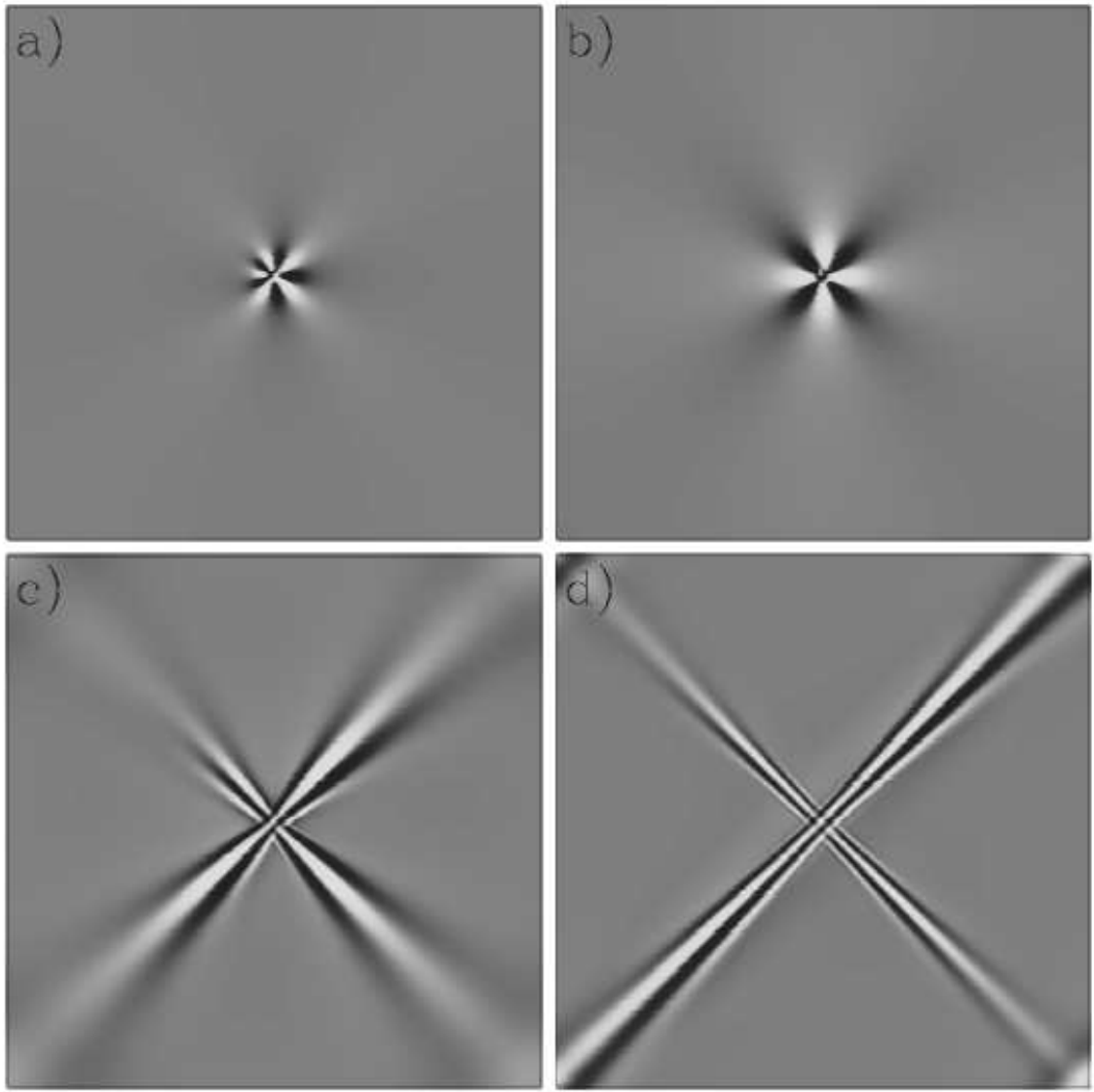}
\caption{The nuclei of Figure 1 at a later time. 
}
\end{figure}

\pagebreak

.

\pagebreak

\begin{figure}[t]
\epsfysize=6.3in
\epsfbox[-20 00 620 580]{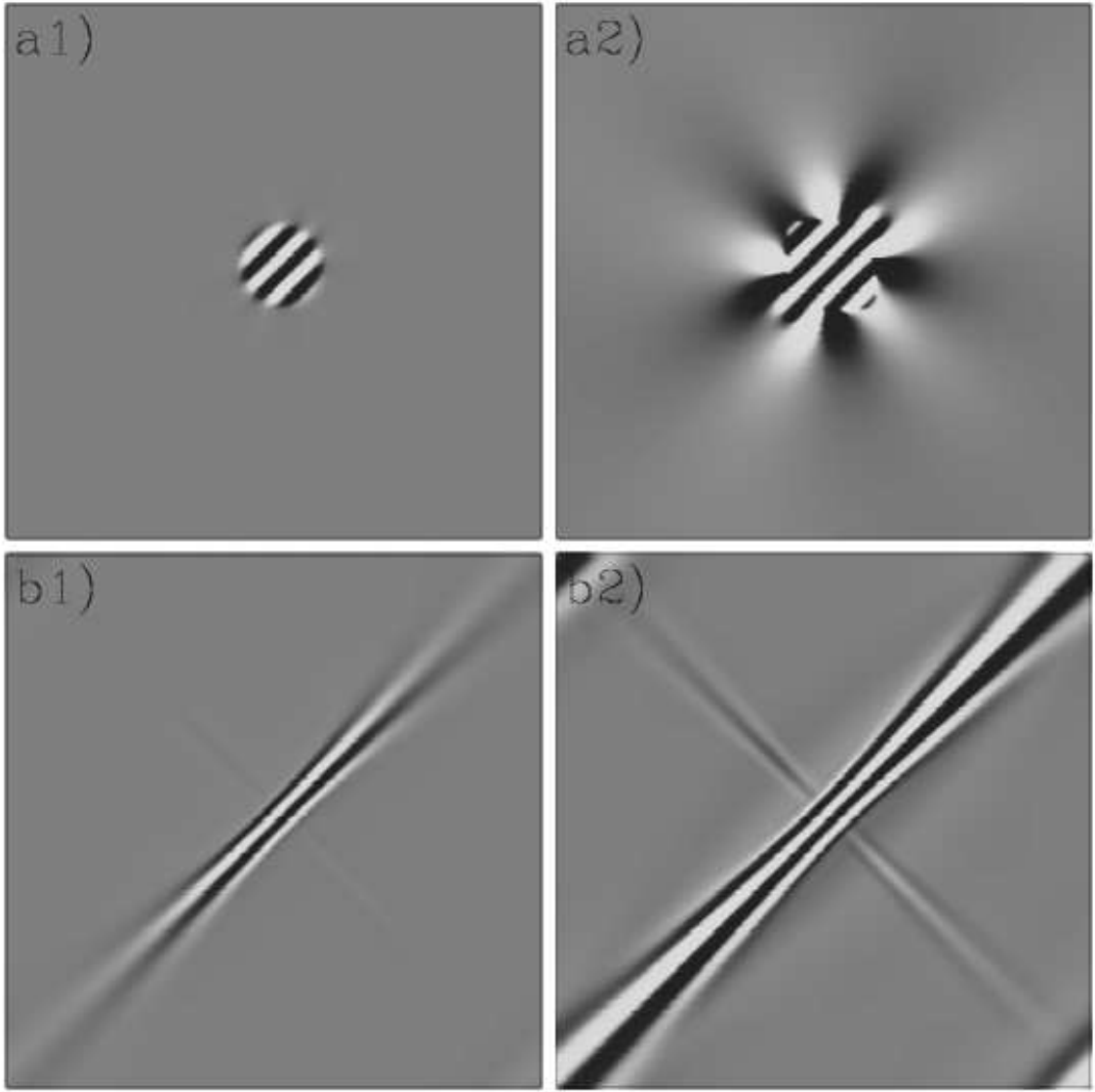}
\caption{Snapshots of O nuclei growing in a T matrix. 
The starting configuration was a disk, internally twinned. 
Parts (a1) and (a2) show the nucleus at times $0.12$ and $0.24$ for a soft
system ($A_1'=0$, $A_2'=1$, $A_3'=2$, $\zeta_1=\zeta_3=1$,
$\tau=-50$).
Parts (b1) and (b2) show the nucleus at times $0.12$ and $0.18$ for a stiff
system ($A_1'=0$, $A_2'=1$, $A_3'=2$, $\zeta_1=\zeta_3=1000$,
$\tau=-50$).
}
\end{figure}

\pagebreak

.

\pagebreak

\begin{figure}[t]
\epsfysize=7.0in
\epsfbox[30 -10 550 540]{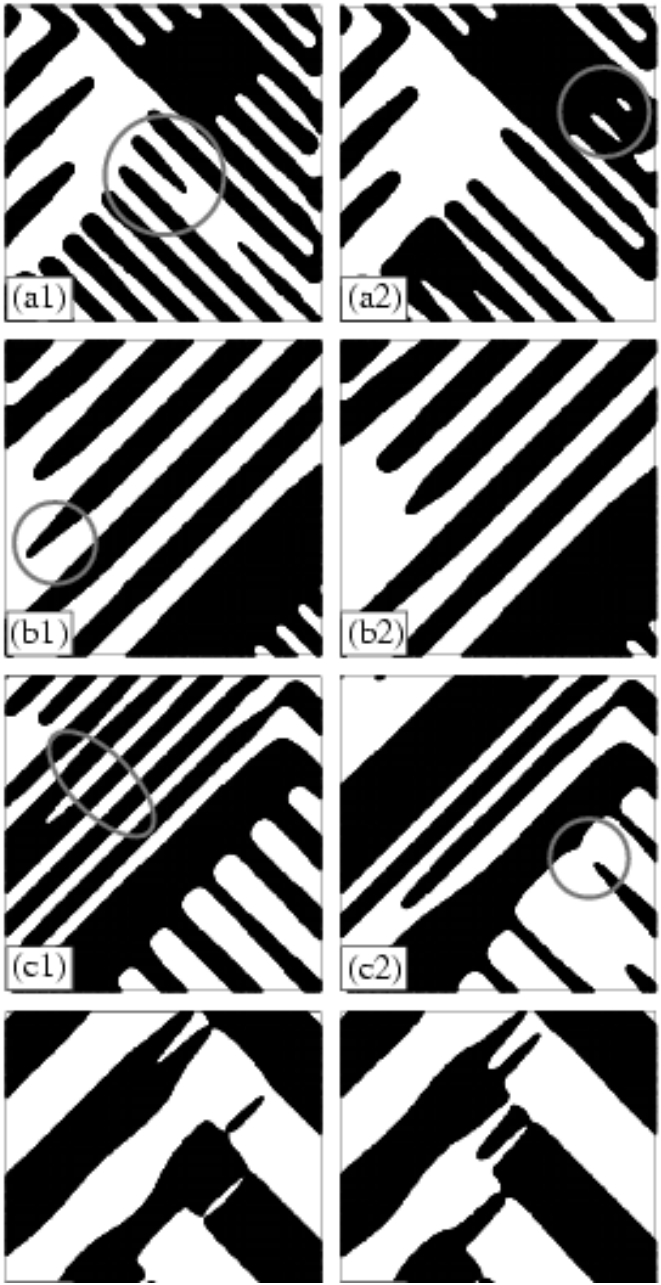}
\caption{Pairs of snapshots showing the time evolution of structures 
for four different initial conditions or parameter sets. 
Each part is a $128\times128$ piece of a full $256\times256$ 
simulation with step size 0.2. 
Parts (a) to (c) are soft systems and part (d) stiff; 
the parameter values and time intervals are given in the text. 
}
\end{figure}

\end{document}